\documentclass[twocolumn,aps,pre,showpacs,floatfix,preprintnumbers,amsmath,amssymb,superscriptaddress,10pt]{revtex4-1}

\usepackage{ulem}
\renewcommand{\emph}{\textit}

\usepackage{selinput}

\usepackage{amsmath}
\usepackage{amssymb}
\usepackage{graphicx}
\usepackage{dcolumn} 
\usepackage{bm} 
\usepackage[usenames,dvipsnames]{color}
\usepackage{natbib}

\usepackage{dsfont}
\usepackage{xr}
\usepackage{mathtools}

\usepackage{pdfpages}

\newcommand{\mean}[1]{\left < #1 \right >}
\newcommand{\abs}[1]{\left | #1 \right |}
\renewcommand{\vec}[1]{\mathbf{ #1 }}

\begin{document}

\title{Superdiffusion, large-scale synchronization and topological defects}

\author{Robert Gro{\ss}mann}
\email{grossmann@physik.hu-berlin.de}
\affiliation{Physikalisch-Technische Bundesanstalt Berlin, Abbestr.~2-12, 10587 Berlin, Germany}

\author{Fernando Peruani}
\affiliation{Laboratoire J.~A.~Dieudonn\'{e}, Universit\'{e} de Nice Sophia Antipolis, UMR 7351 CNRS, Parc Valrose,
F-06108 Nice Cedex 02, France}

\author{Markus B{\"a}r}
\affiliation{Physikalisch-Technische Bundesanstalt Berlin, Abbestr.~2-12, 10587 Berlin, Germany}

\begin{abstract}
We study an ensemble of random walkers carrying internal noisy phase oscillators which are synchronized among the walkers by local
interactions. 
Due to individual mobility, the interaction partners of every walker change randomly, hereby introducing an additional, independent source
of fluctuations, thus constituting the intrinsic nonequilibrium nature of the temporal dynamics.
We employ this paradigmatic model system to discuss how the emergence of order is affected by motion of
individual entities. In particular, we consider both, normal diffusive motion and superdiffusion. 
A non-Hamiltonian field theory including multiplicative noise terms is derived which describes the nonequilibrium dynamics at the
macroscale. 
This theory reveals a defect-mediated transition from incoherence to quasi long-range order for normal diffusion of oscillators in two
dimensions, implying a power-law dependence of all synchronization properties on system size.  
In contrast, superdiffusive transport suppresses the emergence of topological defects, thereby inducing a continuous
synchronization transition to long-range order in two dimensions. 
These results are consistent with particle-based simulations.
\end{abstract}

\date{\today}
%
%

\maketitle

\section{Introduction}
In a large variety of chemical and biological systems, moving entities synchronize an internal
degree of freedom:~examples include mobile catalytic beads exhibiting the oscillatory
Belousov-Zhabotinsky~(BZ) reaction~\cite{Taylor_dynamical_2009} as well as the BZ reaction in 
vortical flow~\cite{paoletti_synchronization_2006}, 
synthetic genetic oscillators~\cite{danino_synchronized_2010},
motile cells synchronizing intracellular oscillations during embryonic
somitogenesis~\cite{jiang_notch_2000,horikawa_noise_2006,Oates_patterning_2012}, 
quorum sensing of signaling amoebae~\cite{gregor_onset_2010}
and the coordinated motion of myxobacteria regulated by the oscillatory
Frz-signaling system~\cite{Sliusarenko_accordion_2006,zhang_mechanistic_2012,starrus_pattern_2012}.
Furthermore, it has been speculated that the flagellar beating of active microswimmers synchronizes
by hydrodynamic interaction thus affecting the rheology of active
suspensions~\cite{furthauer_phase_2013,leoni_synchronization_2014}.

Motivated by the above-mentioned nonequibrium systems, we study synchronization in an ensemble of either diffusively or  
superdiffusively moving random walkers carrying internal noisy oscillators which synchronize by local interactions.  
This paradigmatic model describes a nonequilibrium system as evidenced by an analogy to Monte-Carlo
algorithms~\cite{deMasi_rigorous_1985}:~random motion corresponds to Kawasaki
dynamics~\cite{kawasaki_diffusion_1966} at infinite temperature, whereas the oscillator dynamics in the presence of noise
is similar to Glauber updates~\cite{glauber_time_1963} at a finite temperature. 
Ergo, the system is coupled to two heat baths simultaneously and, as a consequence, does never reach thermodynamic
equilibrium.

We are primarily addressing the question how the emergence of order in spatially extended systems, which relies on the
suppression
of fluctuations by sufficiently fast information transfer through the system, is influenced by the
motion type of individual entities. 
This question is of interest beyond synchronization in nonequilibrium statistical mechanics, ranging from 
 active matter~\cite{romanczuk_active_2012,marchetti_hydrodynamics_2013,menzel_tuned_2015} and 
collective
motion~\cite{toner_flocks_1998,Vicsek_collective_2012} to stochastic reaction-diffusion
systems~\cite{hinrichsen_non_2000,tauber_applications_2005} and epidemic
spreading~\cite{frasca_dynamical_2006,peruani_dynamics_2008,Hallatschek_acceleration_2014}.  
Ultimately, the interplay of local interaction and transport resulting in the emergence of order is a key element in 
all the above-named applications. 

The study of phase transitions and spontaneous symmetry breaking in isothermal systems at thermodynamic equilibrium has already led to
fundamental theoretical insights regarding the emergence of order such as the universality of critical behavior~\cite{hohenberg_theory_1977}
or the Mermin-Wagner theorem~\cite{mermin_absence_1966}, which excludes the emergence of long-range order in low-dimensional equilibrium
systems with a continuous symmetry at finite temperature. 
This work focusses on the crucial influence of the motility of individual entities on the emergence of macroscopic order in
nonequilibrium systems, which is less well understood. 
%
%

%
The temporal dynamics of many oscillators models, such as phase oscillators which constitute one of the corner stones of synchronization
theory~\cite{pikovsky_universal_2001}, is invariant under global phase shifts.  
%
%
This leads to the emergence of spin-waves in one dimension~\cite{peruani_mobility_2010,uriu_dynamics_2013} 
and topological defects~\cite{chaikin_principles_2000} in two dimensions. 
These facts, that relate synchronization theory and statistical mechanics, were often overlooked in
studies of motile oscillators~\cite{porfiri_random_2006,Skufca_communication_2004,frasca_synchronization_2008,fujiwara_synchronization_2011,
gomez_motion_2013,Uriu_random_2010,uriu_optimal_2012}.  
Here, we discuss how the motion of these topological excitations and the corresponding annihilation dynamics is affected by the motion of
the oscillators.  

Whereas various previous studies focused on deterministic
oscillators~\cite{porfiri_random_2006,Skufca_communication_2004,frasca_synchronization_2008,fujiwara_synchronization_2011,
gomez_motion_2013}, we examine fluctuation-induced properties of synchronized states such as correlation
functions and study synchronization from a statistical mechanics perspective.
%
%
In particular, we systematically derive a field theory from the oscillator model  
%
and demonstrate that oscillators moving diffusively exhibit a
defect-mediated transition from incoherence to quasi long-range order (QLRO) analogous to the
Berezinskii-Kosterlitz-Thouless
(BKT) transition~\cite{berezinskii_destruction_1971,kosterlitz_ordering_1973} in two dimensions.
We show further that a continuous nonequilibrium transition to long-range order (LRO) does occur~--~forbidden in
equilibrium systems according to the Mermin-Wagner theorem~\cite{mermin_absence_1966}~--~if oscillators move 
superdiffusively, thereby accelerating the annihilation of defects. 

\section{Model}
We consider an ensemble of $N$ random walkers in a periodic space, each of them carrying an internal clock. Individual clocks, 
which are assumed to possess identical mean natural frequencies $\omega_0$, are modeled as noisy phase oscillators 
that synchronize to the clocks of neighboring walkers. 
The state of the $j$-th walker is thus characterized by its position $\vec{r}_j(t)$ and the phase
$\theta_j(t)$ or the relative phase $\chi_j(t) = \theta_j(t) - \omega_0 t$. 
The dynamics of $\vec{r}_j(t)$ and $\chi_j(t)$ obeys the Langevin equations 
 \begin{subequations}
 \label{eqn:model}
 \begin{align}
\!  \dot{\vec{r}}_j & \!= \boldsymbol{\xi}_{j}(t), \label{eqn:model:a} \\
\!\!\!  \dot{\chi}_j    &\!=\! \frac{1}{N_j} \! \sum_{k\neq j} \beta \! \left ( \abs{\vec{r}_{k} - \vec{r}_j} \right)
\sin \left
(\chi_k -\chi_j \right) \!+ \! \sqrt{2 D_{\chi}} \eta_j(t) \label{eqn:model:b}\! \, . \!\!\!\!\!
 \end{align}  
 \end{subequations}
This model system can be understood as a nonequilibrium, off-lattice version of the XY model 
where individual spins are motile.

The random motion of particles (or spins) is described by Eq.~\eqref{eqn:model:a}, where $\boldsymbol{\xi}_{j}(t)$
denotes L\'{e}vy noise~\cite{klages_anomalous_2008,NoteGaussianNoise}. 
Hence, a particle jumps according to $\vec{r}_j(t\!+\!\tau) = \vec{r}_j(t) \!+\! \int_t^{t+\tau} d t'
\,\boldsymbol{\xi}_{j}(t')$ in a time interval $\tau>0$. 
The random displacements follow a L\'{e}vy $\alpha$-stable distribution~\cite{samoradnitsky_stable_1994} defined by its
propagator~$G_\alpha(\vec{r},\tau)$ in Fourier domain:~$ \hat{G}_{\alpha} \! 
\left (\vec{q} ,\tau \right) = e^{- \gamma \tau \,\! \abs{\vec{q}}^{\alpha}} \!$.
The motion is characterized by a generalized diffusion coefficient $\gamma$ determining the width of the
displacement distribution, and an index $\alpha$ describing the tails of the step length distribution. 
For~$\alpha = 2$, the displacement distribution is Gaussian and, consequently, particles
undergo Brownian motion with a diffusion coefficient $\gamma$. 
In contrast, the step length distribution is heavy tailed for $\alpha \! \in \! (0,2)$ implying large jumps.
Hence, walkers perform L\'{e}vy flights~\cite{klages_anomalous_2008,Metzler_the_2001}.  

Eq.~\eqref{eqn:model:b} describes the oscillator dynamics. The first term accounts for the
local interaction among oscillators analogous to the paradigmatic Kuramoto
model~\cite{Kuramoto_cooperative_1984,shinomoto_cooperative_1986,pikovsky_universal_2001,acebron_kuramoto_2005}.
The interaction strength $\beta$ of two oscillators is a function of their relative distance. 
We assume that the interaction is strictly short-ranged, i.e.~it vanishes beyond a certain interaction
range representing the sensing radius of an individual oscillator that we set to one:~$\beta(x) = 0$ for $\abs{x}>~1$.  
The interaction is rescaled by the number of neighbors $N_j = \sum_{k\neq j} \beta \! \left ( \abs{\vec{r}_{k} - \vec{r}_{j}}
\right)$ reflecting that the adaptivity of an oscillator is independent of the number of neighbors.
The second term in Eq.~\eqref{eqn:model:b} is an additive, Gaussian white noise~\cite{gardiner_stochastic_2010}
with intensity $D_{\chi}$.

\section{Phenomenology} 

\begin{figure}[b]
 \begin{center}
   \includegraphics[width=\columnwidth]{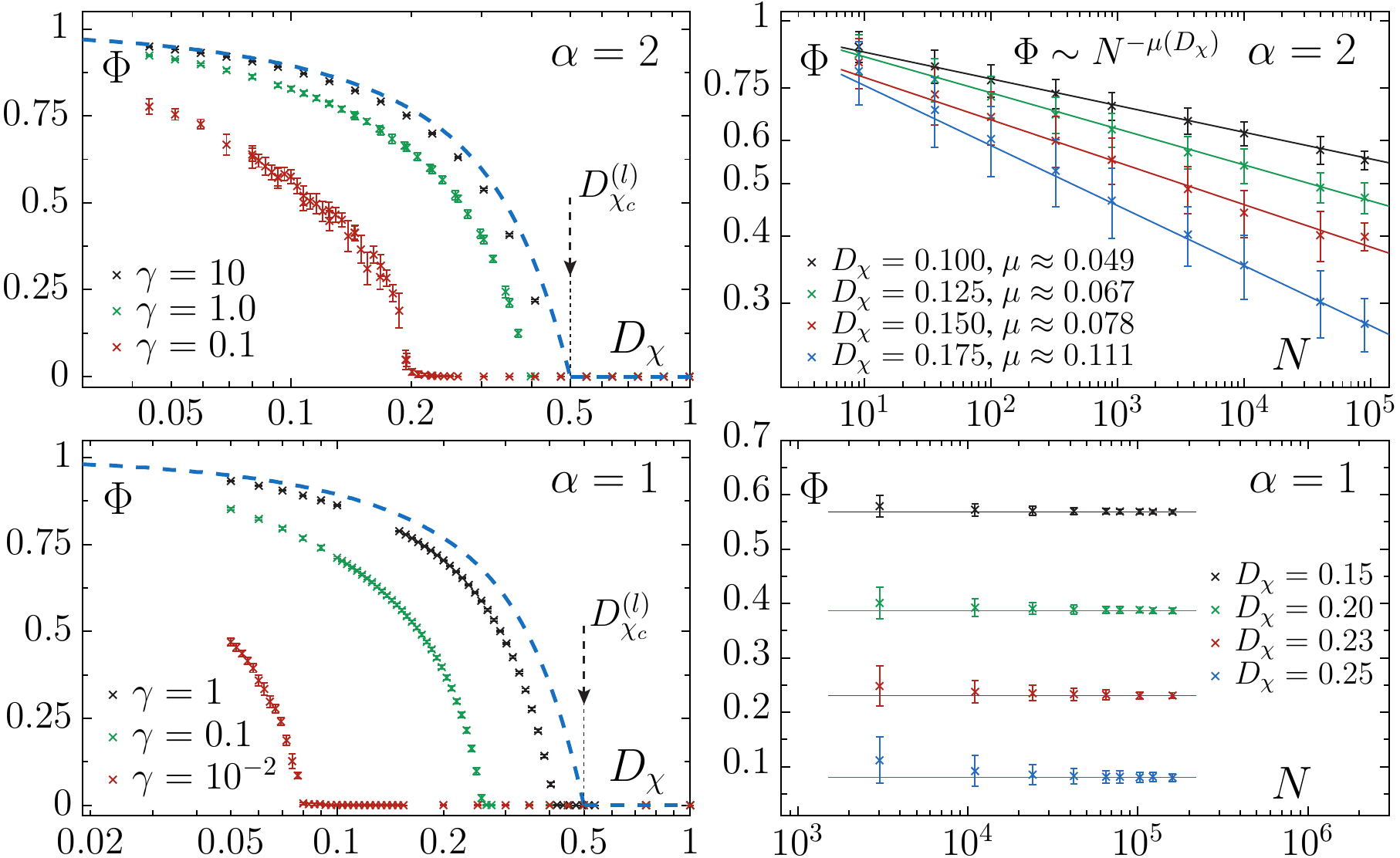}
   \caption{Order parameter $\Phi$ (averaged over time) as a function of the noise $D_\chi$ (left) and the 
particle number $N$ (right) for oscillators moving diffusively (upper row) and by L\'{e}vy flights (lower row). The blue 
dashed line represents the solution of the Kuramoto model (critical point:~$D_{\chi_c}^{(l)} = 1/2$) for global
coupling~\cite{NoteGlobalKuramoto}. In accordance with the theory (cf. section~\ref{sec:fieldTheory}), the lines in the upper right panel
correspond to power law fits, 
while the horizontal lines in the lower right panel indicate the saturation of the order parameter with system size. Parameters: $\rho_0 =
1$, $\beta(x)=\Theta(1-x)$, $L=300$ (left column), $\gamma = 0.1$ (right column), numerical time step $\tau = 0.01$.}
   \label{fig:OrderPar}
 \end{center}
\end{figure}

\begin{figure}[t]
 \begin{center}
\includegraphics[width=\columnwidth]{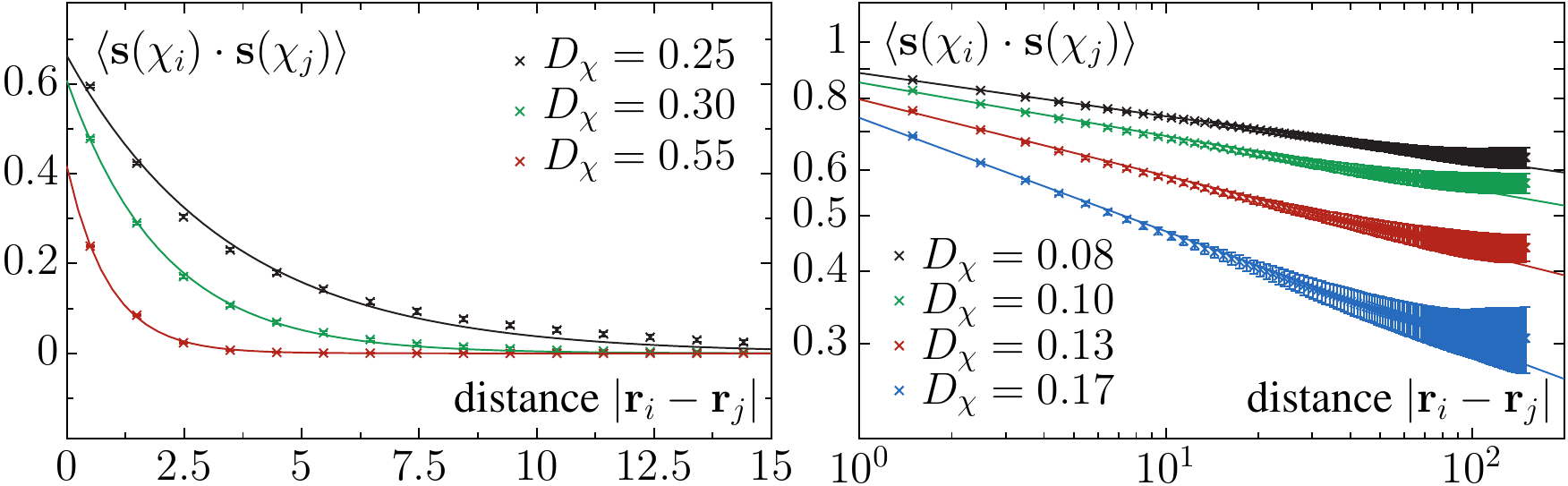}
  \caption{Correlation function (averaged over time) of $\vec{s}(\chi_i) = (\cos \chi_i, \sin \chi_i)$ as a function of
the distance for diffusing oscillators. For high noise (disordered phase), correlations decay exponentially (left panel). In
contrast, correlations decrease algebraically for low noise (right panel) confirming the BKT nature of the synchronization
transition and the emergence of QLRO. Lines represent fits using the associated functional form. Parameters: $\alpha = 2$, $\gamma
= 0.1$, $\rho_0 = 1$, $L = 300$, $\beta(x) = \Theta(1-x)$, numerical time step $\tau = 0.01$. }
  \label{fig:CorrFunc:A2}
 \end{center}
\end{figure}

\begin{figure}[b] 
\begin{center}
  \includegraphics[width=0.85\columnwidth]{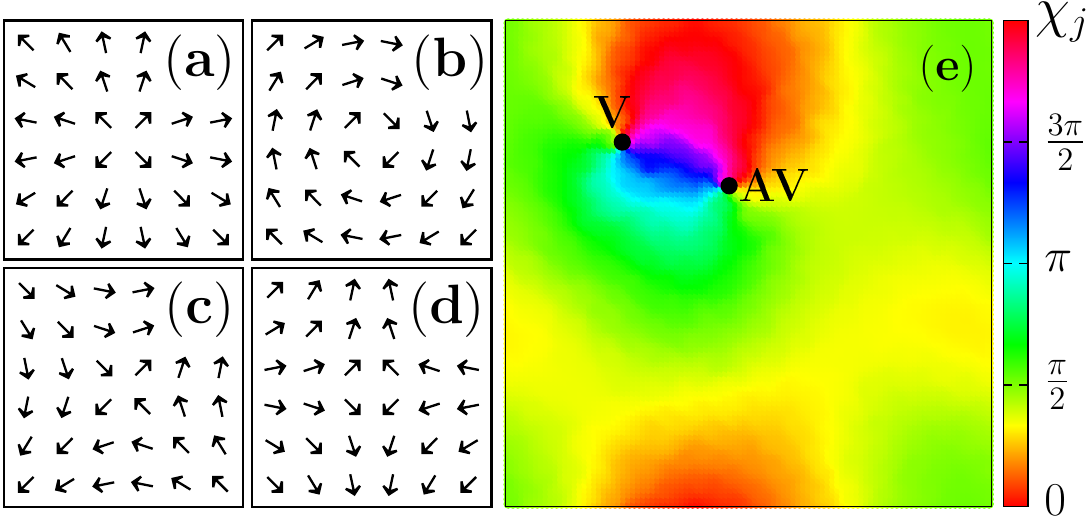}
  \caption{Relaxational dynamics from random initial conditions:~vortices~(V) and anti-vortices~(AV)
are schematically shown by vectors $\vec{s}(\chi_j) = (\cos \chi_j, \sin \chi_j)$ in panels (a,b) and (c,d), respectively. A snapshot of a
vortex-anti-vortex pair is depicted in
(e), where phases~$\chi_j$ are color-coded. } 
  \label{fig:Vortices}
\end{center}
\end{figure}

\begin{figure}[t] 
\begin{center}
  \includegraphics[width=0.95\columnwidth]{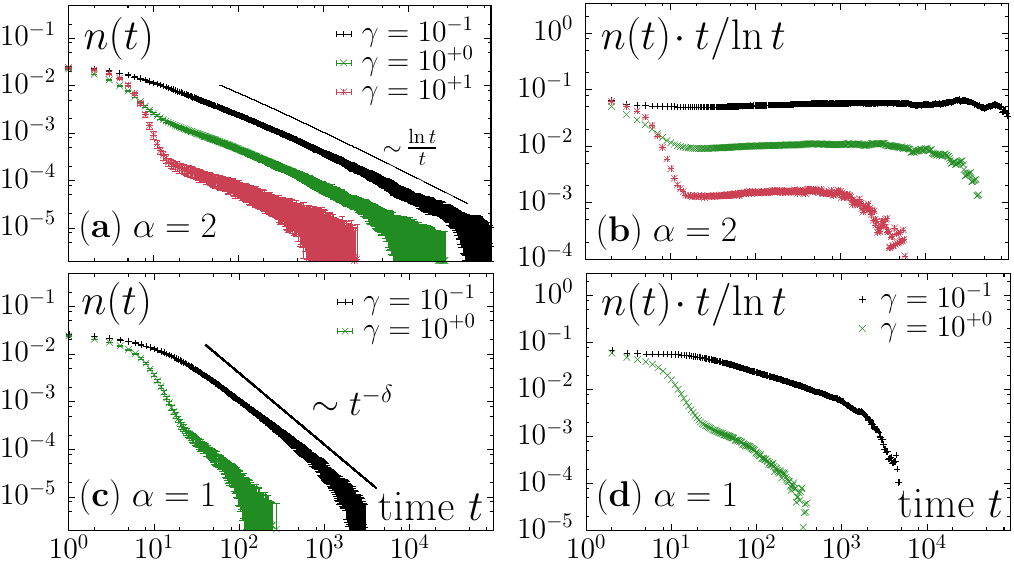}
  \caption{Relaxational dynamics from random initial conditions:~defect density $n(t)$ (averaged over multiple realizations)
for Brownian motion (top) and L\'{e}vy flights (bottom) as a function of time. On the right, the defect density is rescaled
by $t/\!\ln t$. The black line in (a) also indicates this scaling. Superdiffusive transport (d) qualitatively enhances the relaxational
dynamics as compared to the case of normal diffusion. Parameters:~$\rho_0 = 1$, $D_\chi = 0$, $L = 300$ (a,b), $L=400$ (c,d),
$\beta(x)=\Theta(1-x)$, numerical time step $\tau = 0.01$. } 
  \label{fig:RelDyn}
\end{center}
\end{figure}

We performed numerical simulations particularly studying the influence of oscillator motion on synchronization in two dimensions. 
Here, the density of particles is set to $\rho_0 = 1$, which is below the percolation threshold $\rho_{0,perc.} \approx 1.44$~\cite{quintanilla_efficient_2000}. 
%
%
Therefore, the communication of oscillators over large scales is dominated by particle transport.
First, we consider the order parameter $\Phi = \abs{\sum_{k=1}^N e^{i \chi_k(t)} \!/N}^2$ as a function of the
noise strength $D_\chi$ (left column in Fig.~\ref{fig:OrderPar}).  
A transition from incoherence ($\Phi \simeq 0$) to synchronization ($\Phi > 0$) is observed below a critical noise
strength~$D_{\chi_c}$. 
The dependence of $D_{\chi_c}$ on the diffusivity $\gamma$ indicates that synchronization is enhanced by
increasing $\gamma$.  
For $\gamma \rightarrow \infty$, the order parameter tends to the order parameter for globally coupled
oscillators~\cite{NoteGlobalKuramoto}. 
We can conclude that motion enhances information transfer in finite systems, thus promoting the emergence of order.

The transport properties crucially influence the synchronization properties which becomes evident from varying
the particle number $N$ and system size $L$ such that the average density $\rho_0 = N/L^2$ is constant
(finite-size scaling).
For Brownian motion, the order parameter decreases according to a power law with the
particle number (top right in Fig.~\ref{fig:OrderPar}).
The corresponding exponent is a function of the model parameters, particularly the noise strength~$D_{\chi}$. 
Thus, the system is globally disordered in the thermodynamic limit.  
%
%
Similar information is conveyed by the correlation function~(Fig.~\ref{fig:CorrFunc:A2}).  
We assigned to each phase variable~$\chi_i$ a spin $\vec{s}(\chi_i) = (\cos \chi_i, \sin
\chi_i)$ and calculated the correlations of $\vec{s}$ as a function of the
distance. 
Correlations decay exponentially for high noise but decrease according to a power law in the low noise regime as
QLRO emerges through a BKT transition~\cite{berezinskii_destruction_1971,kosterlitz_ordering_1973}.

L\'{e}vy flights change the qualitative picture significantly:~the order parameter $\Phi$
tends towards a nonzero constant value for large system sizes~(bottom right in Fig.~\ref{fig:OrderPar})
suggesting the emergence of a robust synchronized state.
Further, this result implies, notably, the emergence of long-range order (LRO) in two dimensions. 
%
%
%
We find that superdiffusion, in contrast to what we observe for diffusive transport, qualitatively affects the large-scale properties 
of the ordered states and, consequently, the nature of the associated phase transition. 
%
In section \ref{sec:fieldTheory}, we confirm these findings by a field theoretic analysis. 
%

The emergence of long-range order in the case of superdiffusive transport can intuitively be understood by studying in which way macroscopic
fluctuations decay. We quantify the temporal relaxation to the stationary state by preparing the system in the disordered state and, 
thereafter, observing how the system relaxes for zero noise ($D_{\chi}=0$). 
%
%
After an initial transient, the system is locally ordered but globally disordered due to the emergence of
free topological defects~(vortices and anti-vortices) which eventually annihilate pairwise~(see Fig.~\ref{fig:Vortices} for a 
visualization).  
In general, defects annihilate faster if the diffusivity of oscillators is high~(cf.~Fig.~\ref{fig:RelDyn}). 
However, the asymptotic behavior of the time dependent defect density~$n(t)$ does not depend on the diffusion coefficient~$\gamma$ for
normal diffusion~(black line in Fig.~\ref{fig:RelDyn}a):~$n(t) \sim \ln t / t$ similar to the classical XY
model~\cite{bray_breakdown_2000}.
On the other hand, the annihilation process is qualitatively accelerated towards a power-law like decay of the defect density
$n(t)\sim t^{-\delta}$ with $\delta > 1$ in the long-time limit, if particles perform L\'{e}vy flights indicating that superdiffusion
of individual entities induces a novel regime of the relaxation process.   

\section{Field theory}
\label{sec:fieldTheory}

We corroborate the simulation results by deriving and analyzing field equations for the relevant order parameters. 
The instantaneous particle density 
 \begin{align}
  \rho(\vec{r},t) = \mbox{$\sum_{j=1}^N$} \, \delta \! \left ( \vec{r} - \vec{r}_j(t) \right ) 
 \end{align}
is a central quantity of interest, since the presence of density fluctuations distinguishes the nonequilibrium temporal dynamics
of motile oscillators from lattice systems in equilibrium~--~the dynamic  network of motile particles induces a
fluctuating number of neighbors thus providing a source of nonthermal noise. 
In the stationary state, the mean density is constant since there are no forces acting among the particles thus excluding the 
emergence of density instabilities.  
Fluctuations around the mean are determined by the propagator~$G_{\alpha}(\vec{r},t)$ which also reflects the transition
probability for a spatial displacement of a particle within a certain time
interval: 
 \begin{subequations}
 \label{eqn:PropDensField}
 \begin{align}
   & \!\! \mean{\rho(\vec{r},t)}  = \rho_0 \label{eqn:PropDensFielda} \\
   & \!\! \mean{\rho(\vec{r},t) \rho(\vec{r}'\!,t')} - \rho_0^2 = \rho_0 \! \left [ G_{\alpha}(\vec{r}-\vec{r}'\!,t-t') - L^{-d}
\!\;\right ] \!. \label{eqn:PropDensFieldc} \!\!
\end{align}  
 \end{subequations}
Correlations of the density field are proportional to the mean density and, therefore, the variance of the rescaled density
$\delta \tilde{\rho} = \left [ \rho/\rho_0 - 1 \right ]$ decreases proportional to $\rho_0^{-1}$ in accordance with the central
limit theorem~\cite{feller_introduction_1970}.

Along with the density describing the spatial distribution of particles, we introduce an order parameter field
\begin{align}
 \label{eqn:def:OrderPar}
 \vec{S}(\vec{r},t)  = \abs{\vec{S}} \! \begin{pmatrix}
                          \cos \psi \\ \sin \psi
                        \end{pmatrix} \!
                       = \sum_{j=1}^N \begin{pmatrix}
                          \cos \chi_j(t) \\ \sin \chi_j(t)
                        \end{pmatrix}  \! \delta \! \left ( \vec{r} - \vec{r}_j(t) \right )
\end{align}
reflecting the local degree of synchronization. 
This field variable is related to the global order parameter $\Phi$ that was used to characterize particle 
based simulations via 
\begin{align*}
 \Phi \!=\! \mean{ \abs{\frac{1}{N} \!\int \!\! d^d r\, \vec{S}(\vec{r},t)}^2 } \! = \! \frac{1}{N^2} \!\!\int \!\! d^d r d^d r'
\!\mean{\vec{S}(\vec{r},t) \! \cdot \! \vec{S}(\vec{r}',t)} \!. 
\end{align*}
Hence, an increase of the global order parameter~--~as observed in numerical simulations~--~is related to a qualitative change of
the correlation function of $\vec{S}$:~correlations become long-ranged at the critical point. 
The definition of the order parameter, Eq.~\eqref{eqn:def:OrderPar}, constitutes a change of variables from~$\left \{
\vec{r}_j(t),\chi_j(t) \right \}$ to the field~$\vec{S}(\vec{r},t)$. 
Therefore, the dynamics is obtained by application of stochastic
calculus~\cite{dean_langevin_1996,gardiner_stochastic_2010}. 
It turns out that the dynamics is coupled to the density and the nematic tensor~$\mathcal{Q} =
\sum_{j=1}^N \mathcal{M}[\chi_j(t)] \, \delta(\vec{r} - \vec{r}_j(t))$, where  
\begin{align}
 \mathcal{M}[\varphi] = \begin{pmatrix}
                       \cos \left (2 \varphi \right )  & \; \; \; \;\;\,\sin \left (2 \varphi\right ) \\
\sin \left (2 \varphi\right ) & \; \;- \cos \left (2 \varphi\right )
                      \end{pmatrix} \! .
\end{align}
Here, we use the closure relation $\mathcal{Q} \! \approx \! \abs{\vec{S}}^2 \! M[\psi]/\rho$ to obtain a self-consistent
field equation for $\vec{S}(\vec{r},t)$: 
\begin{align}
 \partial_t \vec{S}(\vec{r},t) \approx \, & \gamma \Delta^{\alpha/2} \, \vec{S} + \tilde{\beta} \Delta 
\vec{S} + \! \left [ \textcolor{white}{\frac{A}{B}} \!\!\!\!\!\! D^{(l)}_{\chi_c} - D_\chi \! - \epsilon \!\,
\abs{\vec{S}/\rho}^{2} \,\! \right ] \! \vec{S}\nonumber \\ 
 & \! + \sqrt{D_\chi \rho(\vec{r},t) /2 } \; \boldsymbol{\Pi} \cdot \boldsymbol{\vec{\eta}}(\vec{r},t). 
\label{eqn:OrderParameterEqn} \!
\end{align}
This field theory reflects the microscopic dynamics:~the
(fractional) Laplacian multiplied by the diffusivity~$\gamma$ describes the transport of
particles~\cite{Metzler_the_2001,klages_anomalous_2008} giving rise to (non-)local coupling; 
the coefficient $\tilde{\beta} \! = \! \left[\int \! d^d r \,
\beta \! \left (\abs{\vec{r}} \right ) \! \abs{\vec{r}}^2\right ]\!/\!\left [ 4d \! \int \! d^d r' \beta \! 
\left(\abs{\vec{r}'} \right ) \right ]$ stems from the interaction among the oscillators; 
the term in brackets indicates the emergence of local order via a pitchfork bifurcation for $D_\chi \! <
 D_{\chi_c}^{(l)} = 1/2$, where $\epsilon = 1/2$. 
Fluctuations on the macroscale are reflected by the Gaussian field~$\boldsymbol{\vec{\eta}}(\vec{r},t)$ in
Eq.~\eqref{eqn:OrderParameterEqn}, which is projected by the matrix 
 \begin{align*}
     \boldsymbol{\Pi} &= \sqrt{1 \!+ \! \sqrt{1\!-\!\abs{\vec{S}/\rho}^{4}}} \, \mathds{1}-
\sqrt{1 \!-\!\sqrt{1\!-\!\abs{\vec{S}/\rho}^{4}}} \, \mathcal{M}(\psi) . 
 \end{align*}
We emphasize (i)~the multiplicative noise comprising the square root of the density and (ii)~the non-Hamiltonian dynamics.
Consequently, the probability to find a certain field configuration is generally not determined by a Boltzmann
distribution, in turn reflecting the absence of temperature and the nonequilibrium nature of the model. 
In this regard, the dynamics of the system close to the transition point may potentially contain novel
physics including a set of non-classical critical exponents in the low density regime. 
These questions which go beyond the scope of this work may be addressed in future studies.  
Here, we focus on the properties of the synchronized phase as well as the emergent 
dynamics of topological defects.  
In the high density regime, fluctuations of the density can approximately be neglected according
to Eq.~\eqref{eqn:PropDensField} enabling us to link the nonequilibrium system featuring a dynamic network of short-ranged
interactions to the field theory of the static XY
model~\cite{fisher_critical_1972,suzuki_wilson_1972,Bergersen_dynamical_1991,zinn_quantum_2002}. 
This nontrivial mapping, which connects two fundamentally different physical regimes, explains the phenomenology observed in
simulations in several aspects. 
We deduce from Eq.~\eqref{eqn:OrderParameterEqn} that the diffusivity $\gamma$ rescales the coupling
coefficient $\tilde{\beta}$ in the case of Brownian diffusion ($\alpha = 2$). 
The structure of Eq.~\eqref{eqn:OrderParameterEqn}, however, does not change for $\gamma \rightarrow
0$ in that case. Therefore, the qualitative phenomenology of the XY model is observed if oscillators perform Brownian motion
including a BKT transition towards QLRO whereas the equation is structurally different for superdiffusive motion.  
Close to the critical point~(in the disordered phase), the local order parameter is small, $\abs{\vec{S}/\rho}\ll 1$, and
therefore~$\boldsymbol{\Pi} \approx 2 \cdot \mathds{1}$ to lowest order. 
In this limit, the field equation~\eqref{eqn:OrderParameterEqn} reduces to the Ginzburg-Landau model with additive white noise
enabling us to determine the qualitative dependence of the transition point
$D_{\chi_c}$ on model parameters~\cite{bellac_quantum_1991,tauber_critical_2014}: 
\begin{align}
 \label{eqn:ShiftDC}
  \!\!D_{\chi_c} \! = \! D_{\chi_c}^{(l)} \! \left [ 1 -  \frac{2 \epsilon}{\rho_0} \! \int^{\Lambda} \! \! \! \!
\frac{d^{\;\! d} \;\!\! q}{(2\pi)^d}
\frac{1}{\gamma \! \abs{\vec{q}}^{\alpha}  \! + \tilde{\beta} \! \abs{\vec{q}}^2} + \mathcal{O}(\epsilon^2) \right ]
\!.
\end{align}
The integral in Eq.~\eqref{eqn:ShiftDC} runs over all wavevectors with~$\abs{\vec{q}} < \Lambda$, where $\Lambda$ is a
cutoff parameter inversely proportional to the coarse-graining scale. 
Since the integrand is positive, the critical noise $D_{\chi_c}$ is shifted towards smaller noise values with
respect to $D_{\chi_c}^{(l)}$ as observed in simulations (cf. Fig.~\ref{fig:OrderPar}).
Only if the mobility or the particle density is high, the noise-induced shift of $D_{\chi_c}$ vanishes and the
order-disorder transitions on the local and global level occur simultaneously:~$D_{\chi_c} \simeq D_{\chi_c}^{(l)}$.  

We focus now on the properties of locally ordered states, where $\vec{S}(\vec{r},t) \!\approx \! S_0 (\cos \psi(\vec{r},t), \sin
\psi(\vec{r},t))$.   
In this regime, all quantities are determined by the phase field~$\psi(\vec{r},t)$, such as the correlation
function~$\mean{\vec{S}(\vec{r},t) \!\cdot\! \vec{S}(\vec{r}',t)} \approx S_0^2 \mean{ \cos \left [ \psi(\vec{r},t) -
\psi(\vec{r}',t) \right ] }$. 
The dynamics of $\psi(\vec{r},t)$ follows from Eq.~\eqref{eqn:OrderParameterEqn}: 
\begin{align*}
 \partial_t \psi \approx 
   \gamma \!\; \mathcal{C} \!\! 
     \int \!\! d^{\;\! d} \;\!\! r' \,\frac{\sin \! \left [
\psi(\vec{r}',t) - \psi(\vec{r},t) \right ]}{ \abs{\vec{r}' -
\vec{r}}^{\alpha + d}} +\!  \tilde{\beta} \Delta \psi \! + \! \sqrt{\varkappa} \, \varsigma .
\end{align*} 
The noise strength is denoted by $\varkappa=D_{\chi}\rho_0(1+S_0^2/\rho_0^2)/S_0^{\!\:2}$, $\mathcal{C}$ is
a constant~\cite{NotePreFact:FractLaplace3} and~$\varsigma(\vec{r},t)$ is a Gaussian white field~\cite{NoteGaussianWhiteField}. 
We assume further that~$\psi(\vec{r},t)$ varies slowly such that we can
approximate $\sin \! \left [ \psi(\vec{r}',t) - \psi(\vec{r},t) \right ] \approx \! \left [ \psi(\vec{r}',t) -
\psi(\vec{r},t) \right ]$ to obtain 
\begin{align}
 \label{eqn:stochastic:ang:eqn2}
 \partial_t \psi(\vec{r},t) \approx \gamma \Delta^{\alpha/2} \psi + \tilde{\beta} \Delta \psi + \sqrt{\varkappa} \,
\varsigma(\vec{r},t). 
\end{align} 
The solution of this linear equation is straightforward via Fourier transformation.
One finds that the stationary phase field $\psi(\vec{r},t)$ is Gaussian allowing us to calculate the order parameter correlations
by the identity~\cite{zinn_quantum_2002} 
\begin{align*}
 \mean{\vec{S}(\vec{r},t) \!\cdot\! \vec{S}(\vec{r}',t)} \! \approx \! S_0^2 \exp \! \left [ - \frac{1}{2} \mean{\left [
\psi(\vec{r},t) - \psi(\vec{r}',t) \right ]^2}\right ] \:\! \!. 
\end{align*}
Accordingly, the order parameter correlation function is determined by the mean squared phase difference which follows from the
solution of Eq.~\eqref{eqn:stochastic:ang:eqn2}: 
 \begin{align}
  \mean{\left [ \psi(\vec{r},t) - \psi(\vec{r}',t) \right ]^2} = \varkappa \! \int^{\Lambda} \! \!\! \frac{d^{\;\! d}
\;\!\! q}{(2 \pi)^d} \,
\frac{1 - e^{i \vec{q} \cdot \left ( \vec{r} - \vec{r}' \right )}}{\gamma \! \abs{\vec{q}}^{\alpha}  \! +
\tilde{\beta} \! \abs{\vec{q}}^2} \, .
 \end{align}
The mean squared phase difference at two points $\vec{r}$
and~$\vec{r}'$ behaves for large separations as $\abs{\vec{r} - \vec{r}'}^{\alpha - d}$, i.e.~it decays for $\alpha
<d$ indicating a long-ranged ordered state due to superdiffusive motion of particles. 
Note that the large scale dynamics essentially depends on the dimension of space. In contrast, microscopic details do not
change the large scale properties qualitatively, thus indicating a form of universality.

\begin{figure}[t] 
\begin{center}
  \includegraphics[width=0.8\columnwidth]{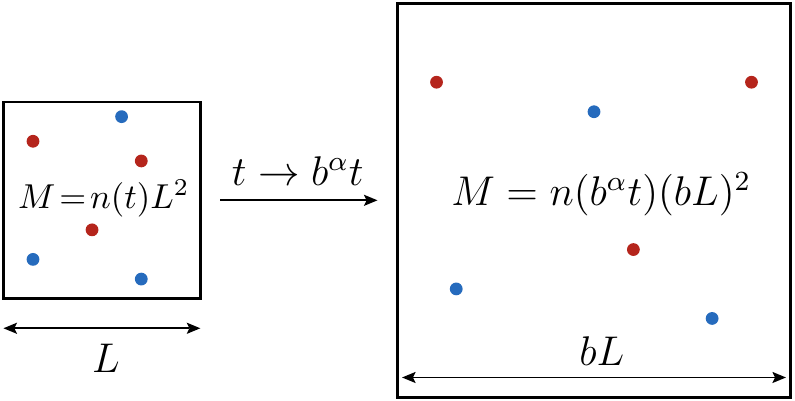}
  \caption{Illustration of the self-similarity argument determining the decay of the density of topological defects. The dynamics of the
phase field $\psi(\vec{r},t)$, cf.~Eq.~\eqref{eqn:stochastic:ang:eqn2}, is asymptotically scale invariant under the rescaling $\vec{r}
\rightarrow b \vec{r}$ and $t \rightarrow b^{\alpha} t$. Therefore, the mean number of defects $M$ in a cube of size $L^2$ at time $t$ (left
panel) and in a cube $(bL)^2$ at time $b^\alpha t$ (right panel) is approximately equal, as illustrated above. } 
  \label{fig:TheoVortices}
\end{center}
\end{figure}

Furthermore, the deterministic equation for $\psi(\vec{r},t)$ supports vortex solutions 
$\psi(\vec{r},t) = \pm \arg(\vec{r}) + \psi_0$ in two dimensions as observed in numerical simulations
(cf.~Fig.~\ref{fig:Vortices}, \ref{fig:RelDyn}). 
We exploit Eq.~\eqref{eqn:stochastic:ang:eqn2} to determine the relaxational dynamics of the defect density from
a qualitative scaling argument. 
By successive rescaling of space $\vec{r} \rightarrow b \vec{r}$ and time $t \rightarrow b^{\alpha} t$ with a
positive scale parameter $b$, the diffusive coupling ($\tilde{\beta} \rightarrow b^{\alpha-2}
\tilde{\beta}$) as well as the noise term ($\varkappa \rightarrow b^{\alpha-2} \varkappa$) are rendered irrelevant for $\alpha
<2$ whereas $\gamma$ remains unchanged. 
Thus, the phase equation becomes scale invariant in the long time limit. 
Hence, the mean defect number in a cube of volume $L^2$ at time $t$ is asymptotically equal to the
defect number in the domain of size $(bL)^2$ at time~$b^{\alpha} t$~(cf. Fig.~\ref{fig:TheoVortices} for an illustration). 
Consequently, the defect density~$n(t)$ is asymptotically a homogeneous function decaying as~$n \! \left
(t \right ) \propto t^{-2/\alpha}$, where the exponent explicitly depends on the motion type via~$\alpha$, but is independent of
microscopic details.
Accordingly, the relaxation is accelerated in the case of superdiffusive transport as observed in numerical simulations
(cf.~Fig.~\ref{fig:RelDyn}).

\section{Summary \& outlook} 
We studied an ensemble of random walkers carrying internal clocks which are synchronized by
local interactions. 
The phenomenology of this paradigmatic nonequilibrium model, as observed in numerical simulations, is remarkably rich. 
Numerical observations were corroborated by a non-Hamiltonian field theory including multiplicative noise terms which was analytically
derived from the microscopic particle dynamics.
The field theory reveals that the qualitative phenomenology of the system is not determined by mixing
rates or, equivalently, the diffusion coefficients but rather by the motion type of particles. 

For Brownian diffusion of individual oscillators, we report on a Berezinskii-Kosterlitz-Thouless transition from incoherence to quasi
long-range order accompanied by the emergence of topological defects implying power-law like system size dependencies of all
synchronization properties.
We argued that the nonequilibrium system with diffusive particle transport exhibits the same
qualitative behavior as an equilibrium system where particles are held fixed on a lattice. 
This implies that long-range order cannot emerge for dimensions smaller or equal to two in the thermodynamic limit and, thus, extends the
predictions of the Mermin-Wagner theorem~\cite{mermin_absence_1966}~--~ valid for equilibrium systems~--~to a nonequilibrium system with
diffusive particle transport. 

Furthermore, we showed that superdiffusive transport may give rise to long-range order
in two and even in one dimension depending on the characteristics of the motion of individual entities.
The emergence of long-range order in two dimensions relies on the accelerated annihilation dynamics of topological defects as confirmed by
numerical simulations in two dimensions. 

In short, we characterized in this work the conditions required for the emergence of long-range order in
nonequilibrium systems. 
We believe that the stochastic dynamics of moving oscillators provides an interesting playground to 
study the interplay of local interaction and particle transport in the context of nonequilibrium
statistical mechanics from a theoretical point of view, involving the mixture of normal fluctuations (oscillator dynamics) and
anomalous fluctuations (superdiffusive motion) as well as the emergence of order and dynamics of topological defects in this context.   
Besides, the results may serve as a reference point for the description of specific experiments such as those mentioned in the
introduction, where modeling may require a higher level of complexity, e.g.~the addition of time delay~\cite{uriu_optimal_2012}.   
  
\section*{Acknowledgement}
 
RG and MB gratefully acknowledge the support by the German Research Foundation via Grant No. GRK 1558. 
FP acknowledges support by the Agence Nationale de la Recherche (ANR) via JCJC project ``BactPhys". 
\appendix 
\bibliographystyle{Myapsrev}
\bibliography{RG_etal_MovingOscillators}

\begin{thebibliography}{61}
\bibitem[{Not({\natexlab{b}})}]{NoteGlobalKuramoto}
\bibinfo{note}{{\textnormal{The order parameter of globally coupled Kuramoto
  oscillators can be obtained from the exact relation $\Phi =
  I_1(\Phi/D_\chi)/I_0(\Phi/D_\chi)$, where $I_n(x)$ denote modified Bessel
  functions of the first kind.}}}
\bibitem[{Not({\natexlab{c}})}]{NotePreFact:FractLaplace3}
\bibinfo{note}{{\textnormal{The prefactor $\mathcal{C} = -2^{\alpha}/\pi^{d/2}
  \cdot\Gamma \! \left ( \frac{\alpha + d}{2} \right )\!/\Gamma \! \left
  (-\frac{\alpha}{2} \right )$ stems from the definition of the fractional
  Laplacian $\Delta^{\alpha/2}$ acting in Fourier space as a multiplication by
  $\left (-\abs{\vec{q}}^{\alpha}\right )$, where $\vec{q}$ denotes the
  wavevector, cf.~\cite{Metzler_the_2001,klages_anomalous_2008}.}}}
\bibitem[{Not({\natexlab{d}})}]{NoteGaussianWhiteField}
\bibinfo{note}{{\textnormal{Besides Gaussianity, Gaussian white fields
  $\xi_{\mu}(x_1,x_2,\dots,x_n)$ are unbiased,
  $\mean{\xi_{\mu}(x_1,x_2,\dots,x_n)} = 0$, and $\delta$-correlated in all
  arguments:
  $\mean{\xi_{\mu}(x_1,x_2,\dots,x_n)\xi_{\nu}(x_1',x_2',\dots,x_n')} =
  \delta_{\mu \nu} \delta(\vec{x} -\vec{x}')$.}}}
\end{thebibliography}

\end{document}